\documentstyle[12pt,aasms4]{article}
\tighten

\def\lsim{\lower.5ex\hbox{$\; \buildrel < \over \sim \;$}}
\def\gsim{\lower.5ex\hbox{$\; \buildrel > \over \sim \;$}}
\def\lax    {\ifmmode{_<\atop^{\sim}}\else{${_<\atop^{\sim}}$}\fi}
\def\gax    {\ifmmode{_>\atop^{\sim}}\else{${_>\atop^{\sim}}$}\fi}
\def\etal{{\it et al.\/} }
\def\gtorder{\mathrel{\raise.3ex\hbox{$>$}\mkern-14mu
             \lower0.6ex\hbox{$\sim$}}}
\def\ltorder{\mathrel{\raise.3ex\hbox{$<$}\mkern-14mu
             \lower0.6ex\hbox{$\sim$}}}

\def\pmb#1{\setbox0=\hbox{#1}%
  \kern-0.015em\copy0\kern-\wd0
  \kern0.03em\copy0\kern-\wd0
  \kern-0.015em\raise0.0433em\box0 }

\begin{document}

\title{ Timing Spectroscopy of Quasi-Periodic Oscillations in the Low-Mass
X-ray Neutron Star Binaries}

\author{Lev Titarchuk}
\affil{NASA/ Goddard Space
Flight Center, Greenbelt MD 20771, and George Mason University/CSI, USA;
titarchuk@lheavx.gsfc.nasa.gov}

\author{Vladimir Osherovich}
\affil{NASA/Goddard Space Flight Center/RITSS, Greenbelt MD 20771 USA;
vladimir@urap.gsfc.nasa.gov}

\author{Sergey Kuznetsov }
\affil{ High-Energy Astrophysics Department, Space Research Institute of
Russian Academy of Sciences, Profsoyuznaya 84/32, Moscow, 117810,
Russia; skuznetsov@hea.iki.rssi.ru}

\vskip 0.5 truecm


\begin{abstract}

Precise simultaneous measurements of the frequencies of the two 
kiloHertz quasi-periodic oscillations (referred in the literature as 
upper and lower kHz QPOs) 
cast doubts on the validity of the simple beat-frequency interpretation 
and some of the modifications introduced to explain the results 
of the varying frequency difference. 
A new model explains the variation of the frequency difference  suggesting 
that the upper kHz QPO, namely $\nu_h$ is an upper hybrid frequency  
of the Keplerian oscillator under the influence of the Coriolis force and 
the lower kHz QPO
is the Keplerian frequency $\nu_K$. Such an oscillator
has two branches characterized by high frequency $\nu_h$ ($\sim$ 1 kHz)
 and by low frequency $\nu_L$ ($\sim 50$ Hz). The frequency $\nu_L$ depends
strongly on the angle $\delta$ between the normal to the neutron star
disk and $\bf{\Omega}$ - the angular velocity of the magnetosphere surrounding
the neutron star. In the lower part of the QPO spectrum ($\sim$ 10 Hz), 
this model identifies the frequency of radial viscous oscillations 
$\nu_{\rm v}$ (previously called \lq\lq extra noise component") and 
the break frequency $\nu_b$, which is associated with the diffusive process 
in the transition region (the innermost part of the disk). 
According to this model, all frequencies (namely $\nu_h$, $\nu_L$, $\nu_b$ 
and $\nu_{\rm v}$) have specific dependences on $\nu_K$. 
This paper focuses on the verification of the
predicted relations. 
For the source 4U 1728-34, the best theoretical fit is
obtained for $\delta=8.3\pm1.0^o$, which is slightly larger than
$\delta=5.5\pm 0.5^o$ previously found for Sco X-1. In addition, we
show that the theoretically derived power law relation 
$\nu_b\propto\nu_{\rm v}^{1.61}$ is consistent with the recent observations 
of other atoll and Z-sources.

\end{abstract}


\section{Introduction}

 KiloHertz quasi-periodic oscillations (QPO) in low mass X-ray neutron star 
(NS) binaries (LMXB's) discovered by 
 Rossi X-Ray Timing Explorer  (Strohmayer \etal 1996, van der Klis
\etal 1996, Zhang \etal 1996) made it possible to evaluate
quantitatively the previously suggested beat-frequency model (Alpar \&
Shaham 1985). Most of the nineteen sources with QPO have twin peaks in
the kHz part of the spectrum (lower frequency marked by K and the higher one
by h in the spectrum of the source Sco X-1 shown in Figure 1).
Within the beat-frequency interpretation, the upper kHz QPO represents 
the Keplerian frequency (van der Klis \etal 1996) while the lower kHz QPO is 
produced by the beating of this frequency with  the NS spin frequency, 
$\nu_{spin}$, which has been identified with the frequency
of nearly coherent oscillations observed during bursts, $\nu_{burst}$  or 
half that (Strohmayer \etal 1996, Miller \etal 1998). According to
Mendez \& van der Klis (1999), \lq\lq if this interpretation is
correct, $\Delta\nu$ equals $\nu_{spin}$ and should remain constant.
The NS spin cannot change by $\sim20\%-30\%$ on time scales of days 
to months." 
For a number of sources 
it has been shown that $\Delta\nu=\nu_h-\nu_K$ decreases systematically
with $\nu_K$ (van der Klis et al. 1997, Psaltis et al. 1998, Markwardt, 
Strohmayer \& Swank, 1999, Mendez et al. 1998, Mendez \& van der Klis 1999).  
These recent findings have undermined the beat-frequency model.
The new (two-oscillator)  model (Osherovich and Titarchuk 1999a and  
Titarchuk and Osherovich 1999, hereafter OT99a and TO99, respectively)  
assumes that the lower frequency $\nu_K$ is the Keplerian frequency.
Following Titarchuk, Lapidus \& Muslimov
(1998 hereafter TLM), we assume a viscous transition layer in the NS disk 
(see Figure 1) which is bounded by the solid surface of the NS  
(the inner boundary) and  by the first Keplerian orbit (the outer
boundary). The  adjustment of the Keplerian disk to the sub-Keplerian inner
boundary  creates conditions favorable for the formation of a hot blob
at the outer boundary of the transition layer (TLM).
When thrown out into the magnetosphere, the blob
becomes a Keplerian oscillator under
the influence of the Coriolis force
 with two modes: a) a radial mode with
frequency
\begin{equation}
\nu_h=[\nu_K^2+(\Omega/\pi)^2]^{1/2}
\end{equation}
where $\Omega$ is the angular velocity of the rotating magnetosphere,
 and b) a mode perpendicular to the disk with frequency
\begin{equation}
\nu_L=(\Omega/\pi)(\nu_K/\nu_h)\sin\delta
\end{equation}
where $\delta$ is the angle between $\bf\Omega$ and the vector normal to
the plane of the Keplerian oscillations ($\delta<<1$). From the observed 
$\nu_h$ and $\nu_K$, using (1), one finds $\Omega$ as a 
function of $\nu_K$ and verifies the existence of the angle $\delta$ which 
satisfies equation (2) for all observed $\nu_K$ and $\nu_h$.
For Sco X-1, it was  found that $\delta=5.5\pm0.5^o$ (OT99a).
The second oscillator in the two-oscillator model has the  frequency of
viscous oscillations $\nu_{\rm v}$ for which the dependence of
$\nu_{\rm v}$ on $\nu_K$ is obtained through the dimensional analysis of
the radial transport of the angular momentum in the transition layer
controlled by a viscous force determined by the Reynold's number, which, in
turn, is related to the mass accretion rate (TLM, TO99). The solution
for $\nu_{\rm v}$ can be presented in the form
\begin{equation}
\nu_{\rm v}=f(\nu_K)C_N
\end{equation}
where $f(\nu_K)$ is a universal function for any NS and the
constant $C_N$ reflects the properties of the specific source.
 The diffusive process in the transition layer is
characterized by a break frequency $\nu_b$ (see Figure 1) which is
related to $\nu_{\rm v}$ by a power law (TO99):
\begin{equation}
\nu_b\propto\nu_{\rm v}^{1.61}.
\end{equation}
For the source 4U 1728-34, the coefficient of proportionality is found
to be 0.041 (TO99). 
The angle $\delta$,
the constant $C_N$ and the profile of $\Omega$ 
determine the frequency range of $\nu_L$, $\nu_{\rm v}$ and $\nu_b$.
Some of the QPOs have been described previously within the 
framework of the two-oscillator model (TO99). For the first time, 
we identify all predicted frequency branches and  
 compare the  inferred main
physical parameters for two stars (section 2, 3 and 4).  
In section 5, we extend the
comparison of the theoretical relation (4) with observed spectra from
other atoll sources and Z-sources. The summary and discussion are
presented in section 6.
\medskip

\centerline{\bf 2. Classification of QPOs in Sco X-1}

\medskip
The frequencies of the observed QPOs for Sco X-1 are plotted in Figure 2
as a function of  $\nu_K$. From the observed $\nu_h$ 
(upper hybrid frequency marked by asterisks in Figure 2) and $\nu_K$, 
the profile of
$\Omega(\nu_K)$ has been calculated and modeled using the theoretically 
inferred magnetic multipole structure of the differentially rotating
magnetosphere (OT99a). The resulting $\Omega$ profile is
\begin{equation}
\Omega/2\pi=C_0+C_1\nu_K^{4/3}+C_2\nu_K^{8/3}+C_3\nu_K^4
\end{equation}
where $C_2=2(C_1C_3)^{1/2}$. The constants
$C_0\equiv\Omega_0/(2\pi)=345$ Hz, $C_1=-3.29\cdot 10^{-2}$ Hz$^{-1/3}$,
$C_2=1.017\cdot10^{-5}$ Hz$^{-5/3}$ and $C_3=-7.76\cdot10^{-10}$
Hz$^{-3}$ have been obtained  by the least-squares fit with $\chi^2=37.6/39$
(OT99a). 
Using equation (5) for $\Omega$, from formula (2) we have calculated $\nu_L$ 
for different $\delta$. The best fit to the data 
presented by open circles in Figure 2 has been found for $\delta=5.5^o$.
  The second harmonic of the low branch,
$2\nu_L\sim90$ Hz observed by van der Klis \etal 1997 
 has been also fitted by our theoretical curve in Figure 2. 
  We employ  the distance determination 
in the function Lebesgue's space, $L_2$ (e.g. Korn \& Korn, 1961, \S 15.2.2)
defining the Relative Lebesgue's  measure $\varepsilon_{RL}$   as a root of 
the ratio of the sum of the square difference (between the theoretical and 
observational values) and the sum of square of the theoretical values. 
Presumably $\varepsilon_{RL}$  is  a better  measure of the difference 
between observations and theory than the $\chi^2$ criteria  in the case 
of the actual large uncertainty in determining the QPOs and break
frequencies. This uncertainty is likely significantly larger than 
statistical errors because the low frequency features are broad and lie atop 
continuum noise of unknown shape.  
If the data set is not uniform for the better control one can use the second 
measure: the rms deviation, $\varepsilon_{rms}$
 which is defined as a ratio of the rms and the mean value of the frequency 
range, (cf. Bendat \& Piersol 1971). Typically 
$\varepsilon_{RL}\leq\varepsilon_{rms}$; for $\nu_L$ in Fig. 2
$\varepsilon_{RL}=\varepsilon_{rms}=$4.6\%.
The frequencies $\nu_L$ and $2\nu_L$
are marked as HBO in the original spectra for Sco X-1 QPOs presented by
Wijnands \& van der Klis 1999 (Figure 3 in their paper).
In the same figure, the break frequency was marked (with question mark) 
and also an extra noise component was shown ($\nu_{\rm v}$ in our 
interpretation).
We have also analyzed the archival data obtained by the RXTE observatory in 
the course of Sco X-1 monitoring on May 25-28, 1996. Standard tasks from 
FTOOLS v4.2 were used in our data analysis according to the RXTE Cook Book. 
To obtain a broad band power density spectra (PDS), the original 1/8192 s 
data were regrouped in 32 s segments with 1/4096 s time resolution data and each
spectrum was calculated over a continuous time interval with a constant
offset. The  PDS obtained were fit over the whole range (1/32 -- 2048 Hz)
using the King and Lorentzian profiles. The Lorentzian components were 
added to the model when their statistical significance was higher than 
$4\sigma$.
 In order to characterize the break in each PDS,
we assume such a \lq\lq break frequency" $\nu_{b}$, 
at which the power drops to 50\% of
 the low frequency plateau value (in terms of the King model).   
Our confidence in the
identification of $\nu_b$ and $\nu_{\rm v}$ (Figs. 1-2) for Sco X-1 is based 
on two tests.
 Firstly, the observations for the two branches in the lower part of the spectra
satisfy our theoretical relation (4), namely, $\nu_b=0.041\nu_{\rm
v}^{1.61}$. As a second test, we use the verification that the extra noise 
component has a frequency which satisfies the theoretically derived 
dependence of $\nu_{\rm v}$ on $\nu_K$. In formula (3), the
universal function $f(\nu_K)$ obtained numerically from a solution for radial
viscous oscillations at the outer edge of the transition region, in
fact, can be approximated by a polynomial
\begin{equation}
\nu_{\rm v}=C_N\cdot(a_1\nu_K+a_2\nu_K^2+a_3\nu_K^3+a_4\nu_K^4)
\end{equation}
where $a_1=1.5033\cdot10^{-3}$, $a_2=1.2289\cdot10^{-6} $Hz$^{-1}$,
$a_3=1.176\cdot10^{-8} $Hz$^{-2}$ and $a_4=1.2289\cdot10^{-11}
$Hz$^{-3}$. The best fit for $\nu_{\rm v}$ ($\varepsilon_{RL}=8.4$\%)
 presented in Figure 2 by a solid 
line (marked viscous) is found for $C_N=9.76$. The line for the related 
$\nu_b$ with the observational points (marked by filled circles) 
closely matching the theory ($\varepsilon_{RL}=13.9$\%) is also shown 
in Figure 2. 
Notice that in vicinity of $\nu_K\sim$800 Hz the line for $\nu_{\rm v}$ 
crosses the line for $\nu_L$ suggesting the possibility of interactions
of two oscillators. 
\medskip
  
\centerline{\bf 3. Classification of the QPOs in the Source 4U 1728-34}

\medskip
The classification of QPOs in the source 4U 1728-34 is presented in
Figure 3. 
The profile of $\Omega/2\pi$ has been modelled using
observations for $\nu_h$ (Mendez \& van der Klis 1999) and formula (5) 
with constants $C_0=382$ Hz,
$C_1=-5.475\cdot10^{-2}$
Hz$^{-1/3}$, $C_2=1.5132\cdot10^{-5}$ Hz$^{-5/3}$ and
$C_3=-1.0492\cdot10^{-9} $Hz$^{-3}$ for $\chi^2=10.5/8$ (OT99a). A
survey of the observed QPO frequencies in the source 4U 1728-34 was
presented  in  Ford \& van der Klis 1998, (hereafter FVK). The 100 Hz
Lorentzian frequency (presented in the Table 1 there) we identify with $2\nu_L$.
The best fit is found for $\delta=8.3\pm1.0^o$ using the extrapolation 
of $\Omega(\nu_K)$ into the range for 
$\nu_K$ below 550 Hz and also observed $\nu_L$  for $\nu_K>800$ Hz.  
The previously identified $\nu_{\rm v}$ (TO99) corresponds to the LF
Lorentzian frequency in the above mentioned Table 1 of FVK
and $\nu_b$ corresponds to the broken PL break frequency  
for three days March 1, February 24 and February 18. The data for
February 16, as explained in FVK and TO99 have to be treated separately (the
observed $4.83\pm0.69$ Hz oscillation does not belong to the $\nu_b$
branch). For this source, the constant $C_N$ is found to be equal to
7.1. In comparison with the Sco X-1 classification (Figure 2), the 
$\nu_{\rm v}$ and $\nu_b$ branches for the source 4U 1728-34 have a much 
wider range and therefore are verified with greater confidence:
$\varepsilon_{RL}=8.5$\% for $\nu_{\rm V}$ and  
$\varepsilon_{RL}=19.5$\% for $\nu_b$.. 
The branches $\nu_L$ and $2\nu_L$ are not observed simultaneously, but 
still match the theoretical curves ($\varepsilon_{RL}=21$\%
for  $\nu_L$).
\medskip

\centerline {\bf 4. Comparison Between QPO Spectra of Sco X-1 and Atoll 
Source 4U 1728-34}

\medskip
Due to the slightly higher $\nu_h$ for source 4U 1728-34, the resulting
$\Omega$ profile has a higher frequency than the $\Omega$ profile for
Sco X-1. For Sco X-1, $\Omega_{max}/2\pi\approx348$ Hz, while for source 4U
1728-34, $\Omega_{max}/2\pi\approx$ 415 Hz. But even more important for 
the values of $\nu_L,$  is the difference in $\delta$ between the two stars. For the
source 4U 1728-34, $\delta=8.3\pm1.0^o$ and the $2\nu_L$ branch in Figure
3 changes from 124 Hz ($\nu_K=353$ Hz) to 182 Hz ($\nu_K=788$ Hz).
The same $2\nu_L$ branch for Sco X-1 stays below 100 Hz for all observed
$\nu_K$ due to the smaller $\delta=5.5\pm0.5^o$ and the smaller $\Omega$
[see formula (2)]. On the other hand, for Sco X-1 the constant
$C_N=9.76$ is greater than $C_N=7.1$ for source 4U 1728-34. Therefore,
the frequency of viscous oscillations $\nu_{\rm v}$ and the related $\nu_b$
are higher for Sco X-1 for the same $\nu_K$, as shown in Figures 2 and 3.
For the same Reynold's number, the greater $C_N$ implies a higher viscosity
(TO99).
\medskip

\centerline{\bf 5. Relation Between the Break Frequencies $\nu_b$ and Viscous
Frequency $\nu_{\rm v}$}

\medskip
The relation between 
$\nu_b$ and $\nu_{\rm v}$ has been approximated by a power law
(4) (TO99). Recent observations of Wijnands \& van der Klis
(1999) make it possible to verify formula (4) for a number of atoll
sources and Z-sources. In Figure 4, the observed $\nu_b$ are plotted vs. 
$\nu_{\rm v}$ on a log-log scale. The open circles correspond to Z-sources and
asterisks represent data for atoll sources. The straight theoretical line 
confirms that the power law $\nu_b\approx 0.041\nu_{\rm v}^{1.61}$ is correct 
for atoll sources. 
When $\nu_b$ is multiplied by a factor of 4, data for Z-sources 
match the same straight line (closed circles represent $4\nu_b$ for Z-sources in
Figure 4). Thus for Z-sources
\begin{equation}\
\nu_b\approx0.01\nu_{\rm v}^{1.61}.
\end{equation}
It is conceivable that the factor of 4 difference in the proportionality
constant is due to the dominant role of different eigenmodes of the diffusive
process for Z-sources in comparison with atoll sources. If this is the
case, then relation (7) corresponds to the first mode and formula (4)
represents the second mode of break frequency (see e.g. Titarchuk 1994
for the details of the diffusion theory). In general, the diffusion equation 
suggests $\nu_{b}(n)\propto n^2$, where $n$ is the integer number of the 
corresponding eigenmode. 
The  overall $\varepsilon_{RL}$ including the Z-sources 
reduced points (closed circles) is 27\%. 

\medskip

\centerline{\bf 6. Summary and Discussion}

\medskip
 In this {\it Letter} we have identified all predicted frequency branches 
for  two stars. Their frequencies are compared   favorably with 
the two-oscillator model. For the source 4U 1728-34, the angle $\delta$ is found
($\delta=8.3\pm1.0^o$). This is the third star for which $\delta$ has been
inferred from QPO observations. For Sco X-1, $\delta=5.5\pm0.5^o$
(OT99a) and for source 4U 1702-42, $\delta=3.9\pm0.2^o$ 
(Osherovich \& Titarchuk 1999b). 
 The predicted relation (4) has
been confirmed for Sco X-1 and source 4U 1728-34 and also for a number
of other atoll and Z-sources. The difference by a factor of 4 between 
 the value of $\nu_b$ for two groups of stars suggests the participation 
of two modes in the
diffusive process in the viscous transition layer in the disk. By comparison 
Sco X-1 has greater viscosity in the transition layer
than the source 4U1728-34, assuming the same
Reynold's number. We stress the
identification of the lowest of the twin kHz frequencies $\nu_K$ as a Keplerian
frequency in contrast to the previous views. The solutions for $\nu_{\rm
v}$ and $\nu_b$ have been found by integration along the radius $R$. The
conversion of the resulting profiles into the frequency domain is done in
according to the formula for the Keplerian frequency $\nu_K$ (e.g. TO99, Eq.1). 
The same is true for the modelling of the $\Omega$ profile. The resulting
striking correspondence between the observed spectra of QPO (Figures 2
and 3) and the predictions of the two-oscillator model we view as
confirmation of the correct identification.
An extrapolation of $\Omega-$profile is not expected to be accurate 
as the data are insufficient to allow a reliable estimate of
the NS spin frequency. The complexity of the QPO phenomena still leaves 
unsolved problems such as $\sim 1$Hz and $\sim 6$Hz  QPO phenomena. 

We are grateful to Eric Ford, Mariano Mendez, Rudy Wijnands and Michiel van 
der Klis for the data and  Joe Fainberg,  Bob Stone and particularly the 
referee for the fruitful suggestions.

\clearpage

\begin{figure}
\caption{ {\it Top}: Schematic picture illustrating the idea of the two 
oscillator model.
The blob forming at the outer boundary of the transition layer oscillates in
the equatorial plane (h-mode) and also in the plane which is perpendicular
to the disk (L-mode). The arrows with the notations ({\bf K,~ b,~ V}) specify
the sites of the proposed origin of the appropriate frequency. 
For the purpose of illustration we have enlarged the transition layer relative
to the neutron star.
{\it Bottom}: Leahy-normalized power spectrum of the Z-source Sco~X-1 in units
of $F \times P(F), (rms/mean)^2$ (see the similar power spectrum figure in 
Wijnands \& van der Klis 1999). Data presented by solid lines
 were averaged over 4 continuous time
intervals after the observation started on May 25, 1996 at 05:04:13 UTC. The  
total integrated time is $\approx$\ 10 ks. The best-fit model spectrum
(which consists of superposition of King profile and five Lorentzian ones) is
shown by dashed line.     
\label{Fig.1}}
\end{figure}

\begin{figure}
\caption{ Classification of QPO in the Z-source Sco X-1. Solid lines are
theoretical curves.
\label{Fig.2}}
\end{figure}

\begin{figure}
\caption{ Classification of QPO in the atoll source 4U 1728-34.
Solid lines are theoretical curves.
\label{Fig.3}}
\end{figure}

\begin{figure}
\caption{ The break frequency vs the viscous 
frequency. Theoretical curve is presented by solid line.
Observational points (Wijnands \& van der Klis 1999) are presented
by asterisks for atoll and millisecond X-ray pulsar sources (4U 1728-34,
 4U 0614+09, 4U 1608-52, 4U 1735-44, 4U 1812, 4U 1705-44, 1E 1724-3045,
and SAX J1808.4-3658) and by open circles for Z-sources (Cyg X-2, GX
17+2, GX 5-1 GX 340+0, Sco X-1). The solid circles are for
the Z-sources where the break frequency is increased by factor 4
(see text).
\label{Fig.4}}
\end{figure}

\noindent
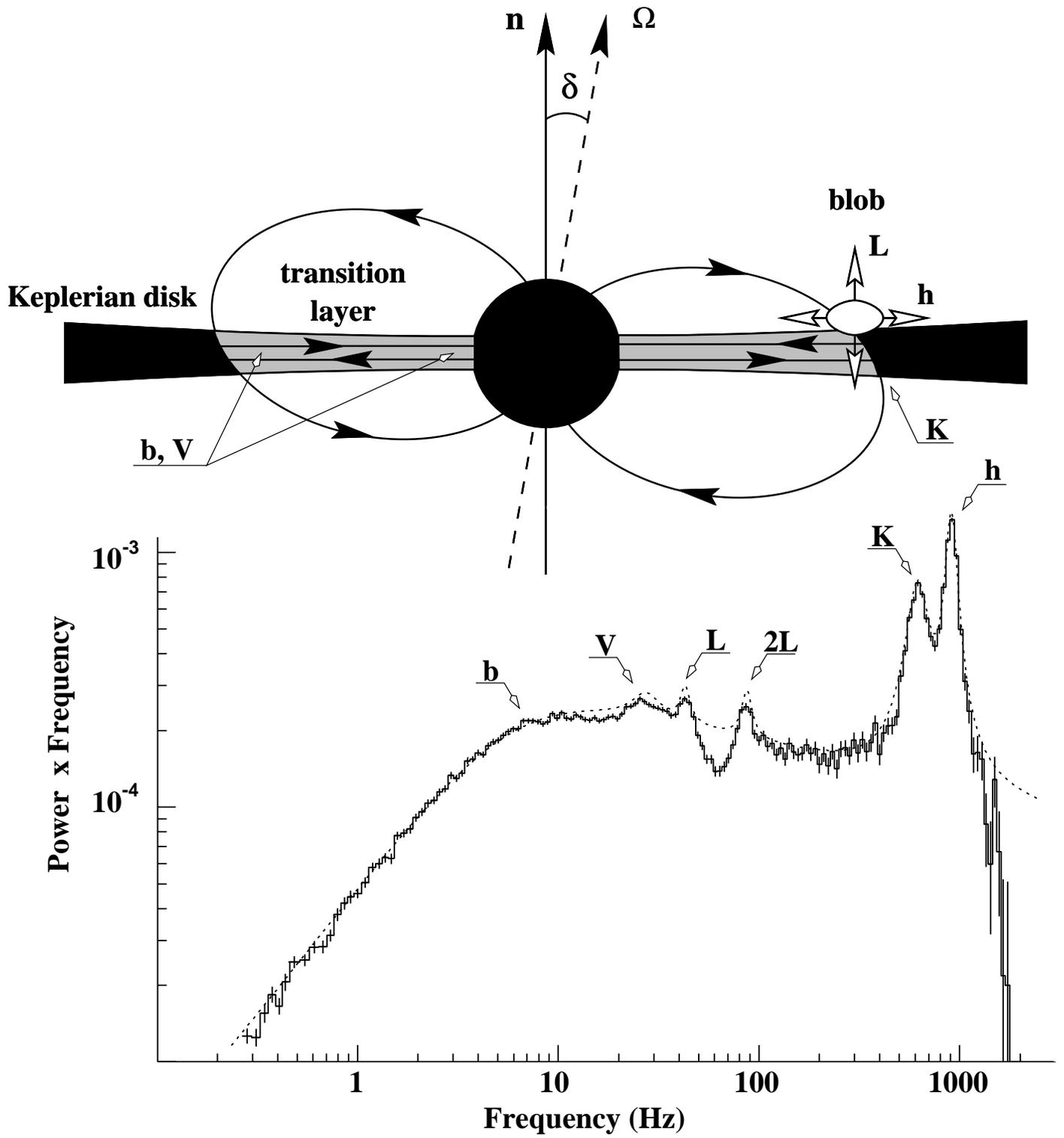
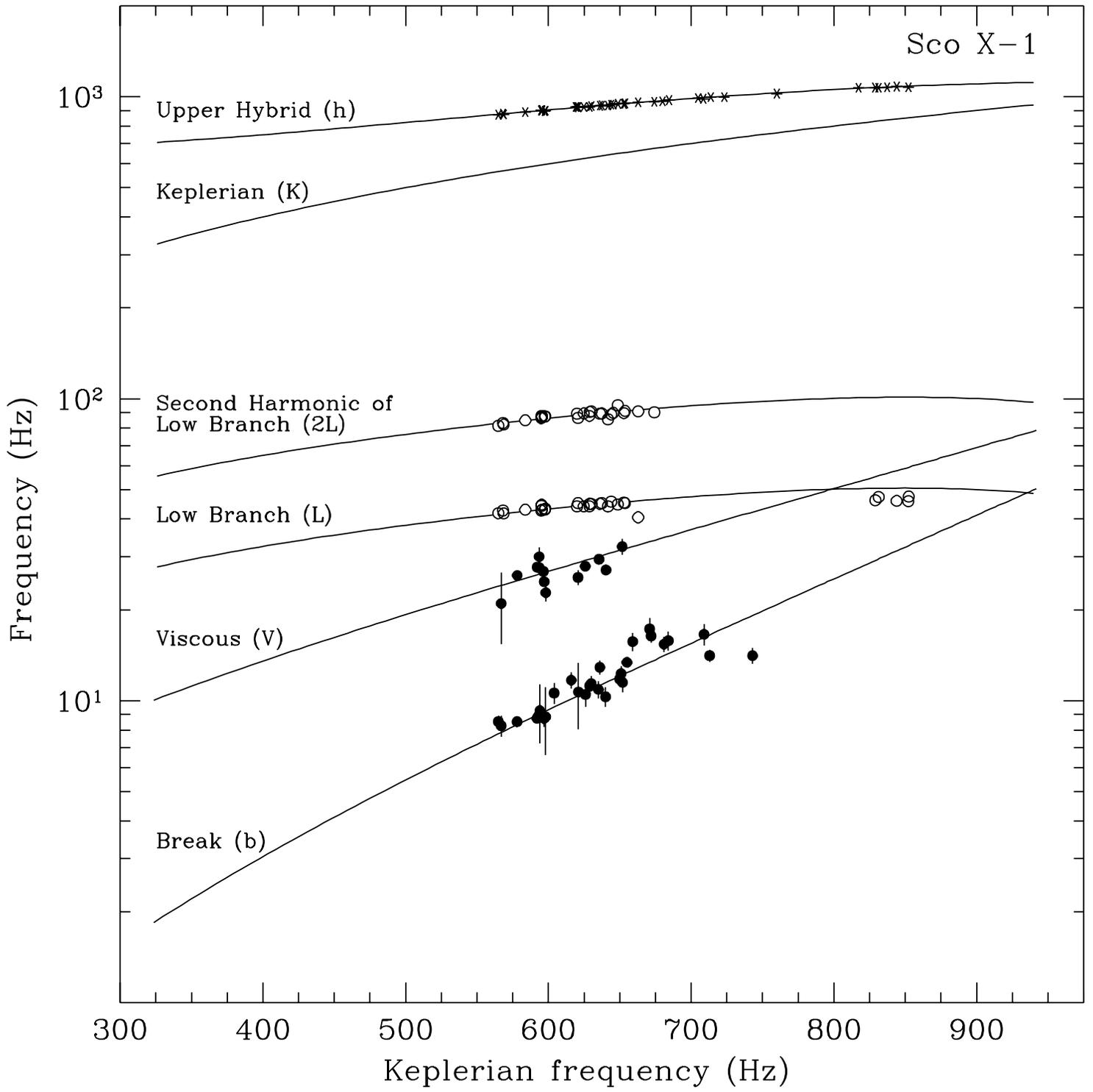
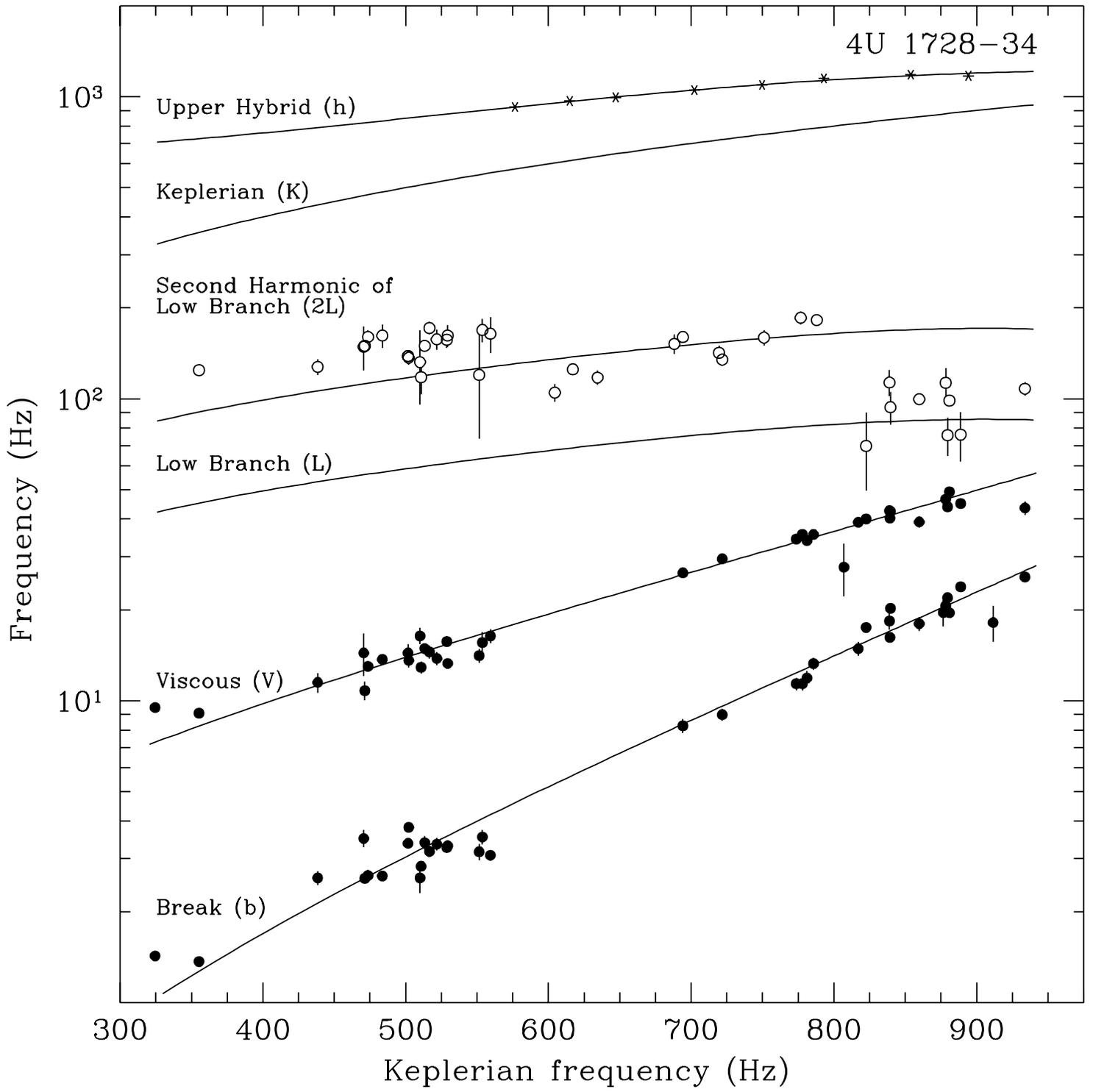
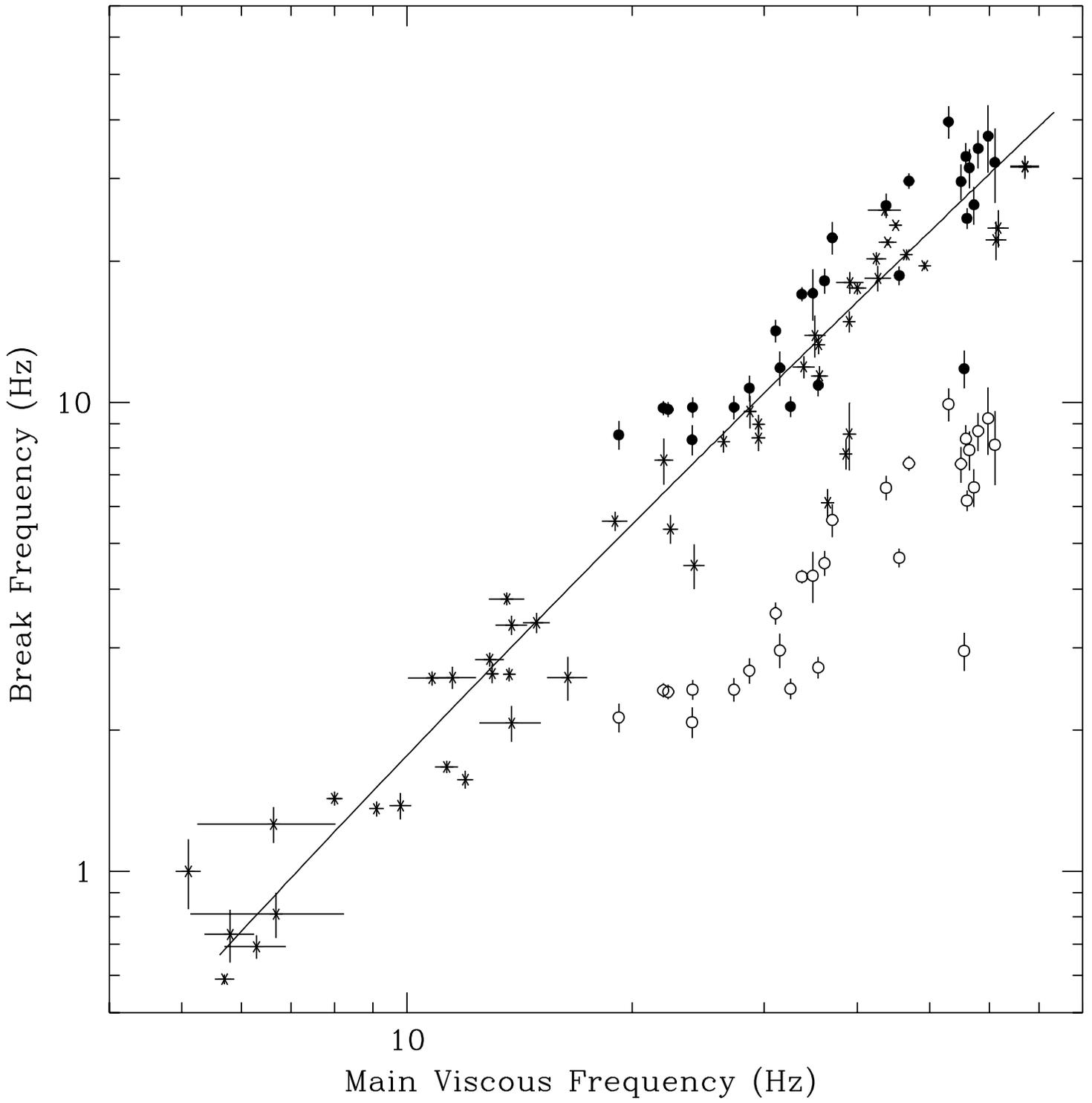
\end{document}